\def \version {2013--12--29}

\documentclass[12pt]{article}
\usepackage[a4paper,left=3.0cm,right=2.5cm,top=3.5cm,bottom=4.5cm]{geometry}

\usepackage{latexsym}
\usepackage{graphicx}
\usepackage{subfigure}
\usepackage{color}
\usepackage{verbatim}

\newcommand{\san}[1]{\mbox{\sf #1}}

\newtheorem{thm}{Theorem}
\def \bt {\begin{thm} \ }
\def \et {\end{thm}}
\newtheorem{crl}{Corollary}
\def \bc {\begin{crl} \ }
\def \ec {\end{crl}}
\newtheorem{cl}{Claim}
\def \bcl {\begin{cl} \ }
\def \ecl {\end{cl}}

\def \qed {\hfill {\boldmath $\Box$}}

\newtheorem{prp}{Proposition}
\def \bpr {\begin{prp} \ }
\def \epr {\end{prp}}
\newtheorem{prm}{Problem}
\def \bpm {\begin{prm} \ }
\def \epm {\end{prm}}
\newtheorem{defi}{Definition}
\def \bdf {\begin{defi} \rm }
\def \edf {\end{defi}}
\newtheorem{lem}{Lemma}
\def \bl {\begin{lem} \ }
\def \el {\end{lem} }
\newtheorem{rem}{Remark}
\def \br {\begin{rem} \ }
\def \er {\end{rem}}
\def \be {\begin{enumerate}}
\def \ee {\end{enumerate}}
\def \bd {\begin{description}}
\def \ed {\end{description}}

\newtheorem{fig}{Figure}
\def \bfg {\begin{fig} \ }
\def \efg {\end{fig}}

\def \cH {{\cal H}}

\def \vp {\varphi}

\def \msk {\medskip}
\def \bsk {\bigskip}

\def \nin {\noindent}
\def \pf {\nin{\bf Proof } }

\newcommand{\G}{G^*}
\newcommand{\astx}{\textasteriskcentered}

\begin{document}

\title
  {~ \vspace{-8ex} \\
  Minimum order of graphs with given\\
    coloring parameters
 \vspace{3mm}}

\author
  { G\'abor Bacs\'o$^1$ \\
    Piotr Borowiecki$^{2,}$\thanks{Research partially supported by National Science Centre under contract DEC-2011/02/A/ST6/00201}\\
    Mih\'aly Hujter$^3$ \\
    Zsolt Tuza$^{4,5}$\thanks{~Research
    supported in part by the Hungarian Scientific Research Fund, OTKA
    grant no.\ 81493, and by the Hungarian State and the European Union
    under the grant TAMOP-4.2.2.A-11/1/ KONV-2012-0072.}\\
     ~~\\
    {\small $^1$~Computer and Automation Research Institute, Hungarian Academy of Sciences,\vspace{-1mm}}\\
    {\small H--1111 Budapest, Kende u.\ 13--17, Hungary\vspace{1mm}}\\
    {\small $^2$~Department of Algorithms and System Modeling,\vspace{-1mm}}\\
    {\small Faculty of Electronics, Telecommunications and Informatics,\vspace{-1mm}}\\
    {\small Gda\'nsk University of Technology, Narutowicza 11/12, 80-233 Gda\'nsk, Poland\vspace{1mm}}\\
    {\small $^3$~Budapest University of Technology and Economics,\vspace{-1mm}} \\
    {\small M\H uegyetem rakpart 3--9, Budapest, Hungary\vspace{1mm}} \\
    {\small $^4$~Department of Computer Science and Systems Technology,\vspace{-1mm}}\\
    {\small University of Pannonia,\vspace{-1mm}}\\
    {\small H--8200 Veszpr\'em, Egyetem u.\ 10, Hungary\vspace{2mm}}\\
    {\small $^5$~Alfr\'ed R\'enyi Institute of Mathematics,\vspace{-1mm}}\\
    {\small Hungarian Academy of Sciences,\vspace{-1mm}}\\
    {\small H--1053 Budapest, Re\'altanoda u.\ 13--15, Hungary\vspace{2mm}}
  }
  
\date{\scriptsize Latest update on \version}

\maketitle

\vspace{-4mm}

\begin{abstract}
A complete $k$-coloring of a graph $G=(V,E)$ is an assignment
 $\vp:V\to\{1,\dots,k\}$ of colors to the vertices such that
 no two vertices of the same color are adjacent, and the
 union of any two color classes contains at least one edge.
Three extensively investigated graph invariants related to
 complete colorings are the minimum and maximum number of colors
 in a complete coloring (\emph{chromatic number} $\chi(G)$ and
 \emph{achromatic number} $\psi(G)$, respectively), and the
 \emph{Grundy number} $\Gamma(G)$ defined as the largest $k$
 admitting a complete coloring $\vp$ with exactly $k$ colors
 such that every vertex $v\in V$ of color $\vp(v)$ has a
 neighbor of color $i$ for all $1\le i<\vp(v)$.
 The inequality chain $\chi(G)\le \Gamma(G)\le \psi(G)$
 obviously holds for all graphs $G$.
 A triple $(f,g,h)$ of positive integers at least 2
 is called \emph{realizable} if there exists
 a \emph{connected} graph $G$ with $\chi(G)=f$, $\Gamma(G)=g$,
 and $\psi(G)=h$. In \cite{ChaOk}, the list of realizable triples
 has been found. In this paper we determine the minimum number of
 vertices in a connected graph with chromatic number $f$, Grundy
 number $g$, and achromatic number $h$, for all realizable
 triples $(f,g,h)$ of integers.
 Furthermore, for $f=g=3$ we describe the (two) extremal graphs
 for each $h \geq 6$. For $h=4$ and $5$, there are more extremal
 graphs, their description is contained as well.

\end{abstract}


  \msk\nin


\section{Introduction}

A \emph{complete coloring} of a graph is an assignment of colors to the
vertices in such a way that adjacent vertices receive different
colors, and there is at least one edge between any two color
classes. In other words, the coloring is proper and the number of
colors cannot be decreased by identifying two colors.

Let $G=(V,E)$ be any simple undirected graph. The minimum number of
colors in a \emph{proper} coloring is the \emph{chromatic number}
$\chi(G)$, and all proper $\chi$-colorings are necessarily complete.
The maximum number of colors in a \emph{complete} coloring is the
\emph{achromatic number} $\psi(G)$. Every graph admits a complete
coloring with exactly $k$ colors for all $\chi\le k\le \psi$ (Harary
\emph{et al.}, \cite{HHP}). An important variant of complete
coloring, called \emph{Grundy coloring} or Grundy numbering,
requires a proper coloring $\vp:V\to\{1,\dots,k\}$ such that every
vertex $v\in V$ has a neighbor of color $i$ for each $1\le
i<\vp(v)$. The largest integer $k$ for which there exists a Grundy
coloring of $G$ is denoted by $\Gamma(G)$ and is called the
\emph{Grundy number} of $G$. Certainly, $\Gamma(G)$ is sandwiched
between $\chi(G)$ and $\psi(G)$. One should emphasize that $\chi(G)$
and $\psi(G)$ are defined in terms of unordered colorings, i.e.,
permutation of colors does not change the required property of a
coloring. On the other hand, in a Grundy coloring the order of
colors is substantial.

Proper colorings have found a huge amount of applications and hence,
besides their high importance in graph theory, they are very well
motivated from the practical side, too. The chromatic number occurs
in lots of optimization problems. The achromatic number looks less
practically motivated, nevertheless it expresses the worst case of a
coloring algorithm which creates a proper color partition of a graph
in an arbitrary way and then applies the improvement heuristic of
identifying two colors as long as no monochromatic edge is created.
Grundy colorings have strong motivation from game theory; moreover,
$\Gamma(G)$ describes the worst case of First-Fit coloring
algorithm when applied to a graph $G$ if we do not know the graph in
advance, the vertices arrive one by one, and we irrevocably assign the smallest
feasible color to each new vertex as a best local choice. Then the
number of colors required for a worst input order is exactly~$\Gamma(G)$. 
For this reason, $\Gamma(G)$ is also called the
\emph{on-line First-Fit chromatic number} of $G$ in the literature.

An overview of on-line colorings and a detailed analysis of the
First-Fit version is given in \cite{online}. A more extensive survey
on the subject can be found in \cite{Kie98}. The performance of
First-Fit is much better on the average than in the worst case. This
is a good reason that it has numerous successful applications. This
nicely shows from a practical point of view that the Grundy number
is worth investigating.

The definition of Grundy number is usually attributed to Christen and Selkow
\cite{ChrS79}, although its roots date back to the works of Grundy
\cite{Gru} four decades earlier; and in fact $\Gamma(G)$ of an
undirected graph $G$ is equal to that of the digraph in which each
edge of $G$ is replaced with two oppositely oriented arcs. In
general, computing the Grundy number is \san{NP}-hard, and it
remains so even when restricted to some very particular graph
classes, e.g., to bipartite graphs or complements of bipartite
graphs (\cite{HS10} and \cite{cobip}, respectively). Actually, the
situation is even worse: there does not exist any polynomial-time
approximation scheme to estimate $\Gamma(G)$ unless $\san{P} =
\san{NP}$ \cite{Kor}, and for every integer $c$ it is
$\san{coNP}$-complete to decide whether $\Gamma(G)\le c\, \chi(G)$,
and also whether $\Gamma(G)\le c\, \omega(G)$, where $\omega(G)$
denotes the clique number of $G$ (see \cite{AHL}).
Several bounds on $\Gamma(G)$ in terms of other graph invariants were given, e.g., in \cite{BoRa13,Zak,Zak2}.
 On the other hand, by the finite basis theorem of Gy\'arf\'as \emph{et al.} \cite{GyKi+97}
 the problem of deciding whether $\Gamma(G)\geq k$ can be solved in polynomial time,
 when $k$ is a fixed integer (see also \cite{BoSi12}  for results on Grundy critical graphs).
Moreover, there are known efficient algorithms to determine the Grundy number of trees \cite{HHB}
and more generally of partial $k$-trees \cite{TP}.

Concerning the achromatic number, on the positive side there exists
a constant-approximation for trees \cite{Chau} and a polynomial-time
exact algorithm for complements of trees \cite{YaGa}. But in a sense,
the computation of $\psi(G)$ is harder than that of $\Gamma(G)$. It
is $\san{NP}$-complete to determine $\psi(G)$ on connected graphs
that are simultaneously interval graphs and co-graphs \cite{Bod},
and even on trees \cite{CaiEd,ManMcD}. Moreover, no randomized
polynomial-time algorithm can generate with high probability a
complete coloring with $C \psi(G)/\sqrt{n}$ colors for arbitrarily
large constant $C$, unless $\san{NP} \subseteq \san{RTime}
(n^{\san{\scriptsize\rm poly}\log n})$, and under the same
assumption $\psi(G)$ cannot be approximated deterministically within
a multiplicative $\lg^{1/4-\varepsilon} n$, for any $\varepsilon >0$
\cite{KorShe}, although some $o(n)$-approximations are known \cite{Chau,KorKra}.

The strong negative results above concerning algorithmic complexity
also mean a natural limitation on structural dependencies, for all
the three graph invariants $\chi, \Gamma, \psi$. On the other hand,
quantitatively, the triple $(\chi,\Gamma,\psi)$ can take any
non-decreasing sequence of integers at least 2. (The analogous
assertion for $(\chi,\psi)$ without $\Gamma$ appeared in \cite{Bh}.)
For example, if $\chi=\Gamma=2$, then 
properly choosing the size of a union of complete graphs on two vertices
will do for any given $\psi$. Assuming connectivity, however,
makes a difference. Let us call a triple $(f,g,h)$ of integers with
$2\le f\le g\le h$ \emph{realizable} if there exists a
\emph{connected} graph $G$ such that $\chi(G)=f$, $\Gamma(G)=g$, and
$\psi(G)=h$. It~was proved by Chartrand \emph{et al.}\ \cite{ChaOk}
that a triple is realizable if and only if either $g\ge 3$ or
$f=g=h=2$.

Here we address the naturally arising question of smallest connected graphs
with the given coloring parameters. Namely, for a realizable triple
$(f,g,h)$, let us denote by $n(f,g,h)$ the minimum order of a
connected graph $G$ with $\chi(G)=f$, $\Gamma(G)=g$ and
$\psi(G)=h$. The lower bound $n(f,g,h)\ge 2h-f$ was proved in
\cite[Theorem 2.10]{ChaOk} in the stronger form $2\psi-\omega$
(where $\omega$ denotes clique number), and this estimate was also
shown to be tight for $g=h$. On the other hand the order of graphs
constructed there to verify that the triple $(f,g,h)$ is realizable
was rather large, and had a high growth rate. In particular,
for every fixed $f$ and $g$, the number of vertices in the graphs of
\cite{ChaOk} realizing $(f,g,h)$ grows with $h^2$ as $h$ gets large,
while the lower bound is linear in $h$. For instance, the
construction for $f=g$, described in \cite{HM}, takes the complete graph $K_f$ together with a pendant path $P_k$, having
properly chosen number of vertices $k$, and applies the facts that very long paths make
$\psi$ arbitrarily large and that the removal of the endvertex
of a path (or actually any vertex of any graph) decreases $\psi$ by
at most 1, as proved in \cite{GK}.

In this paper we determine the exact value of $n(f,g,h)$ for every
realizable triple $(f,g,h)$, showing that the lower bound $2h-f$ is either
tight or just one below optimum. It is easy to see that the complete
graph $K_f$ verifies $n(f,f,f)=f$ for all $f\ge 2$.
For the other cases it will turn out that the formula depends on
whether $f<g$. These facts are summarized in the following
two theorems; the case $g=h$ was already discussed in \cite{ChaOk}.

\bt \label{fkisebbg} For\/ $2 \leq f<g$ and for\/ $f=g=h$,\/
$n(f,g,h)=2h-f$. \et

\bt \label{f=g} For\/ $2<f=g<h$,\/ $n(f,g,h)=2h-f+1$. \et

\br As one can see, the minimum does not depend on\/ $g$, apart
from the distinction between\/ $f=g$ and\/ $f<g$. \er

The two theorems above will be proved in the following two sections,
while in 
the last section for each $h>3$ we determine the number of extremal graphs
(the graphs of minimum order) that realize the triple $(3,3,h)$.


\bt  \label{csak2} Let $f=g=3$. If $h=4$, then there are seven extremal graphs.
If $h=5$, then there are three such graphs, while for every $h\ge 6$ there are exactly two of them.
\et

 Along the proof of Theorem \ref{csak2} we will also determine the structure of all extremal graphs.

 In what follows we need some additional terms and notation.
 Given a graph $G=(V,E)$ and a vertex set $X\subseteq V$,
 the subgraph induced by $X$ will be denoted by $G[X]$.
 A set of vertices $X$ is \emph{dominated by} another one, say $D$, if every $x \in X\setminus D$
 has at least one neighbor in $D$. A set $D$ is \emph{dominating} if it dominates the whole vertex set.
 We will also say that a set $D$ dominates a subgraph $H$ in the sense that $D$ dominates $V(H)$.
 A set $S$ is said to be \emph{stable} if it does not contain any pair of adjacent vertices.

\section{The case $\chi<\Gamma$ --- Proof of Theorem \ref{fkisebbg}}
   \label{c<G}

First, we consider the lower bound for $n(f,g,h)$.
From \cite[Theorem 2.10]{ChaOk}, it follows that for all $f,g,h$ satisfying conditions of Theorem \ref{fkisebbg},
it holds that $n(f,g,h)\geq 2h-f$. The idea
is that $h$ color classes on fewer than $2h-f$ vertices would
yield at least $f+1$ singleton classes, but then they should be
mutually adjacent if the coloring is complete. This leads to the
contradiction $f+1\le \omega(G)\le \chi(G)$.
\msk

Considering the tightness of the bound we construct an appropriate bipartite graph to prove the following lemma

\bl   \label{=2h-f}
 The lower bound\/ $2h-f$ is tight for the case\/ $f=2$. \el

 \nin
{\bf Construction I} \msk

\begin{figure}[h]
\begin{center}
  \includegraphics[width=10cm,clip=true]{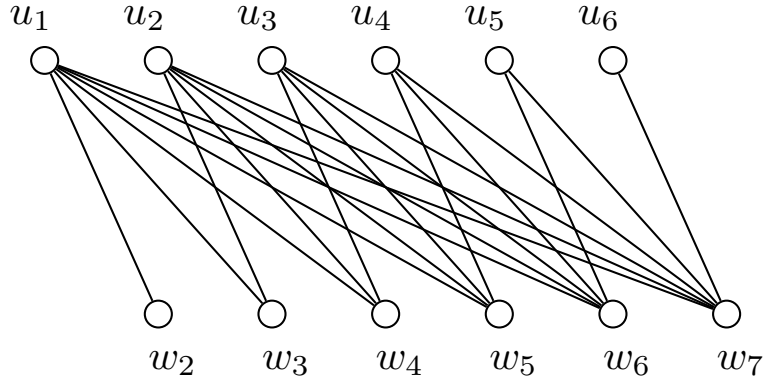}
  \caption{Graph $B_7$}\label{ordered}
\end{center}
\end{figure}

For every $k\ge 2$ we define a \emph{basic bipartite graph} $B_k$,
with vertex set $X=U \cup W$,  where
  $U=\{ u_1,u_2, \dots, u_{k-1} \}$ and
  $W=\{ w_2,w_3, \dots, w_k \}$, and edge set
  $$C=\{ u_i w_j  \mid  1 \leq i < j \leq k \}.$$
We call $U$ and $W$ the \emph{partite sets} of $B_k$.
See Figure \ref{ordered} for the graph $B_7$ that will be used several times.


For $g=3$, we consider $B_h$ itself. However, for $g \geq 4$, we denote
$\gamma=g-3$, and modify the graph $B_h$ by inserting the
following set of edges:
  $$\{ u_i w_j \mid 1 \leq i-j \leq \gamma,
   \ 2 \leq i \leq h-1, \ 2 \leq j \leq h-2 \}.$$
We shall call them the \emph{inserted edges}. \qed

\begin{figure}[h]
\begin{center}
  \includegraphics[width=10cm,clip=true]{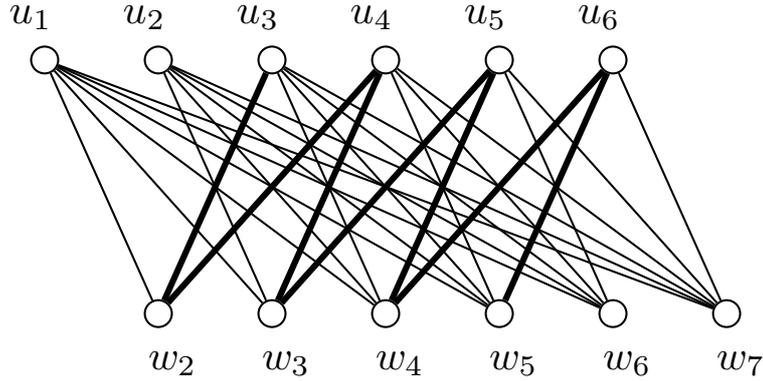}
  \caption{Graph $G(2,5,7)$}\label{comb.}
\end{center}
\end{figure}

\bsk
For any $g \geq 3$, the graph with inserted edges will be denoted by $G(2,g,h)$, and in
the following text, briefly by $\G$. We show an example of $\G$ in Figure
\ref{comb.}.


 \bpr \label{harom}
$\chi(\G)=2$,\/ $\Gamma(\G)=g$, and\/ $\psi(\G)=h$. \epr

 \nin \pf
By definition, $\G$ is bipartite, i.e., $\chi(\G)=2$ and the
color classes $\{u_1\}$, $\{u_2,w_2\},$ $\dots,$
$\{u_{h-1},w_{h-1}\}$, $\{w_h\}$ verify that $\psi(\G)\ge h$ is
valid, whereas $\psi(\G)\le h$ also holds because $\G$ has no more
than $2h-2$ vertices. Hence, what remains to prove is that
$\Gamma(\G)=g$.

For $g<h$, the proof of the lower bound $\Gamma(\G) \geq g$ is obtained
by assigning color $g$ to $u_1$, color $i-1$ to $\{u_i,w_i\}$ for
$i=2,\dots,g-1$, color $1$ again to $u_g,\dots,u_{h-1}$, and finally
color $g-1$ to $w_j$ for $j=g,\dots,h$. Consequently, $\Gamma(\G)
\geq g$.

For $g=h$, the only difference is that in the set $U$, color $1$ is assigned to
exactly one vertex, namely to $u_2$.

To prove the upper bound on $\Gamma(\G)$ is more difficult.
We manage it as a separate statement.

 \bcl \label{gamma} $\Gamma(\G) \leq g$ \ecl

 Throughout the argumentation, we assume that $\Gamma (\G) >g$.
 Claim \ref{gamma} will be a consequence of Claim \ref{BIPnored} below.
 Before proving those claims we need some additional notation and a simple Claim \ref{minmax}.

\msk
Let us call a stable set $S$ a \emph{double set} if it meets both
partite sets. Considering a \emph{Grundy coloring} of $\G$ with
$\Gamma(\G)$ colors, a \emph{double class} is a color class which
is a double set. Graph $\G$ is bipartite and there must be an edge
between any two color classes. Moreover, the classes containing
$u_1$, $w_h$ respectively, are non-double. Thus, the number of
non-double classes is exactly $2$. We denote the double classes by
$D_1,D_2,\dots,D_\Delta$, indexed with their colors; here
$\Delta=\Gamma(\G)-2$.

\msk
Let $S$ be any double set, let $I=I(S)$ be the minimum of $\{i \mid u_i
\in S\}$ and let $J=J(S)$ be the maximum of $\{j \mid w_j\in S\}$. For
$S=D_{k}$ we denote $I(S)$ also by $I_{k}$ and similarly $J_{k}$ for
$J(S)$.

\msk
 The following claim is straightforward, so the proof is omitted.

\bcl \label{minmax} $I(S) \geq J(S)$ for any double set $S$ and,
in particular, for any double class.

\qed \ecl

We shall use the term \emph{reduced graph} and notation $R_t$ for
the bipartite graph $K_{t,t}-tK_2$ obtained from the complete bipartite graph $K_{t,t}$ by
omitting a $1$-factor. (This graph was taken in \cite{ChaOk} for
$f=2$ and $g=h>2$.) The essence of the proof is given in the
following claim. Let us recall that $\Delta$ is the number of double
classes.

\bcl \label{BIPnored}
\be
\item[\textup{(i)}] The subgraph\/ $H$ induced by the vertex set\/
   $$
      \{u_I\mid I=I(D), \; D  \mbox{ is a double class} \}\cup  \{w_J \mid J=J(D), \; D \mbox{ is a double class} \}
   $$
   is isomorphic to\/ $R_{\Delta}$.

\item[\textup{(ii)}] The graph\/ $\G$ does not contain any induced subgraph isomorphic to\/ $R_{\Theta}$ with\/ $\Theta \geq \gamma+2$.
\ee
\ecl

\nin
{\bf Proof}
We start with an observation that nothing has been stated concerning the position of
$\{I_k,J_k\}$ in the ``omitted $1$-factor''.

\msk\nin
In order to prove (i) we take two arbitrary double classes $D_k$ and $D_K$ with $k<K$
and consider the vertices $u_i,u_I,w_j,w_J$ where $i=I_k,$
$I=I_K,$ $j=J_k,$ $J=J_K$. We will prove that $u_I w_j \in E(\G)$
and $u_i w_J \in E(\G)$, this will yield the assertion since
$\{I_k, J_k \mid 1 \leq k \leq \Delta \}$ will play the role of
the ``omitted $1$-factor".

Let us consider the first statement. Suppose for a contradiction
that $u_I w_j \notin E(\G)$. From the properties of Grundy
coloring, $u_I$ has some neighbor $w_{\lambda}$ in $D_k$. By the
maximality of $j$, $j \geq \lambda$ holds and $w_j$ is adjacent to
$u_I$ for every $I\neq j$. We may suppose the latter case since
otherwise the proof is done. Similarly, we obtain that $w_J$ has a
neighbor $u_{\mu}$ in $D_k$.

Since all of $w_{j-\gamma},w_{j-\gamma+1},\dots,w_h$ except $w_j$
are adjacent to $u_I$, the non-edge $u_I w_J$ implies
$J<I-\gamma$, and $i\le \mu$ also holds. Hence, the following
chain of inequalities is valid:
 $$ J < I-\gamma = j-\gamma \le i-\gamma \le \mu-\gamma . $$
Consequently, the vertices $u_I$ and $w_j$ are adjacent, as
claimed.

By the central symmetry of $\G$, the relation $u_iw_J \in E(\G)$ is
established in the same way. \bsk

 \noindent
In order to prove (ii) suppose that $\G$ has an induced subgraph $R$
isomorphic to $R_{\Delta}$ with $\Delta \geq \gamma +2$. Let $Z$
be the vertex set of $R$. If $u \in Z \cap U$, $w \in Z \cap W$,
and $uw \notin E(\G)$, then we call $w$ the \emph{match} of $u$
and vice versa. Let us denote by $a=u_M$ the vertex of largest
subscript in $Z \cap U$ and by $b=w_m$ the vertex of smallest
subscript in $Z \cap W$.

First, suppose $ab \in E(\G)$. In case $M < m$, the graph $R$ would
be a complete bipartite graph. So $M \leq m+\gamma$. Let us take
any vertex $u_{\iota}$ in $(Z \cap U) \setminus \{a\}$. We observe
that $m\leq \iota< M$. Indeed,
 the second inequality is trivial; and if the first one
was not true then $u_{\iota}$ would have no match.
 This means that
all the subscripts of the vertices in $Z \cap U$ are in the
interval $[m,M]\subseteq [m,m+\gamma]$ and thus $\Delta \leq
\gamma +1$, proving the assertion if $ab \in E(\G)$.

Second, suppose $ab \notin E(\G)$. Then the match of $a=u_M$ is
$b=w_m$. Surely, for every vertex $u_{\iota}$ in $Z \cap U$ with
$\iota \neq M$, we have $\iota \leq m+\gamma$ because $u_\iota w_m
\in E(\G)$, and $m<\iota$, for otherwise $u_{\iota}$ would have no
match. Consequently, $\Delta \leq \gamma +1$.

\msk
Thus we have proved Claim~\ref{BIPnored}.
\qed

\msk
As a consequence of the above claim, $\Delta \le \gamma+1=g-2$ and
$\Gamma(\G)\le g$ follow. Thus Claim~\ref{gamma} is established; moreover Proposition \ref{harom} and Lemma \ref{=2h-f} are proved.
\hfill{\boldmath $\Box\Box\Box$}

\msk
Note that we have also proved Theorem \ref{fkisebbg} for $f=2$.
It has been
observed in \cite[Proposition 2.8]{ChaOk} that if we take the
join of the current graph with a new vertex, then each of $\chi,\Gamma,\psi$ increases by exactly 1.
Thus, for a given triple $(f,g,h)$ with $f\ge 3$ we can start from
$G(2,g',h')$, where
$$g'=g-f+2, \quad h'=h-f+2$$
and join it with $K_{f-2}$. In this way we obtain a connected graph
that realizes $(f,g,h)$ and has exactly $(2h'-2)+(f-2) = 2h-f$ vertices.
This completes the proof of {Theorem~\ref{fkisebbg}}.
\qed

\section{The case $\chi=\Gamma$ --- Proof of Theorem \ref{f=g}}
 \label{c=G}

Similarly as above, we shall give the proof in two parts. Before
proving the lower bound, we show an auxiliary statement which will
be applied many times.

\bcl \label{Gamma>} Given a graph $G$, suppose there exists an
induced subgraph $H$ of $G$, with $\Gamma(H)\geq k$
 and a stable set, disjoint from $V(H)$ and dominating $H$.
Then the Grundy number of $G$ is strictly larger than $k$. \ecl

 \pf
For both $G$ and $H$, the stable set can get color 1, and the
vertices of $H$ can be colored with numbers one larger than in the
original Grundy numbering of $H$. This induces a subgraph of
Grundy number larger than $k$, and we can use the fact that
$\Gamma$ is monotone with respect to taking induced subgraphs. \qed

\br In most of the applications,\/ $H$ will be a\/ $K_3$ or a\/
$P_4$ and\/ $k$ will be\/ $3$. As another special case, we
shall often find a maximal stable set of\/ $G$, disjoint from some
subgraph\/ $H$ of\/ $G$. \er

Now we are in a position to establish the first part of the
theorem.

\bl \label{fminus1} If\/ $2<f=g<h$, then\/ $n(f,g,h)\ge 2h-f+1$.
 \el

\pf
 Suppose for a contradiction that there exists a graph
$G$ with the given coloring parameters on $n < 2h-f+1$ vertices. By
what has been said at the beginning of Section \ref{c<G}, this
implies $n=2h-f$. From the argument sketched there we also see that
the conditions $n=2h-f$ and $\psi(G)=h>f$ imply that in any complete
$h$-coloring of $G$ there exist $f$ singleton color classes, say
  $S_1=\{y_1\}, S_2=\{y_2\}, \dots, S_f=\{y_f\}$,
inducing a complete subgraph in $G$. Moreover, by the condition
$h>f$, there is at least one further color class, say $S_{f+1}$.

Since the coloring is assumed to be complete, each $y_i$
$(1 \leq i\leq f)$ is adjacent to at least one vertex of $S_{f+1}$. Hence, by
Claim \ref{Gamma>} we obtain $\Gamma(G) >f=g$, a contradiction. Lemma
\ref{fminus1} is established. \qed \bsk

The proof of the following lemma, which establishes the second part of the theorem, will be split into several claims.

\bl \label{ngeq}
 If\/ $2<f=g<h$, then\/ $n(f,g,h)\le 2h-f+1$.
\el

\pf Two graphs will be constructed with the appropriate number of
vertices, for $f=g=3$. This will be enough since the simple
extension adjoining $K_{h-f+3}$ will work like earlier (see end of the proof of Theorem \ref{fkisebbg}).\bsk

First, we recall the simple fact from the proof of \cite[Proposition
2.5]{ChaOk} that a connected graph $G$ has $\Gamma (G)=2$ if and
only if $G$ is a complete bipartite graph. This can be extended also
to disconnected graphs:

\bcl  \label{g=2}
  For a graph\/ $G$,\/ $\Gamma (G)\leq 2$ is equivalent to the
following property:

$(\Pi)$ Each component of\/ $G$ is either an isolated vertex or a complete bipartite graph.

  ~\qed
 \ecl

Note that under the present conditions we have $h\ge 4$. Starting
from the basic bipartite graph $B_{h-2}$ introduced in the previous
section, we are going to construct two graphs $L_1$ and $L_2$ with
larger parameters by inserting vertices and edges, different from
Construction I.
\msk

\nin
 {\bf Construction II} \msk

Let $\ell=h-2$ and consider the bipartite graph
$B_{\ell}=(X,C)$ of Section \ref{c<G}.
First, we extend $B_{\ell}$ with two isolated vertices
$u_{\ell}$ and $w_1$ to obtain the \emph{extended graph} $B_h'$. We also introduce the notation
$U'=\{ u_1, \dots, u_{\ell} \}$, $W'=\{ w_1, \dots, w_{\ell} \}$.
Let $L_i=(Y,D_i)$, $i\in\{1,2\}$ be the graph with the vertex set $Y=U'\cup W'\cup\{q_1,q_2\}$, and the edge set
$D_1=D_0 \cup \{q_2u_{\ell}, q_1w_1, q_1q_2 \} $ or
$D_2=D_0 \cup \{q_2u_{\ell}, q_1q_2 \}$,  respectively, where
$D_0=C\cup \{q_1u \mid u\in U'\}\cup \{q_2w \mid w\in W' \}$.  \qed

\begin{figure}[h]
\begin{center}
\subfigure[Graph $L_1$]{\includegraphics[width=6.8cm,clip=true]{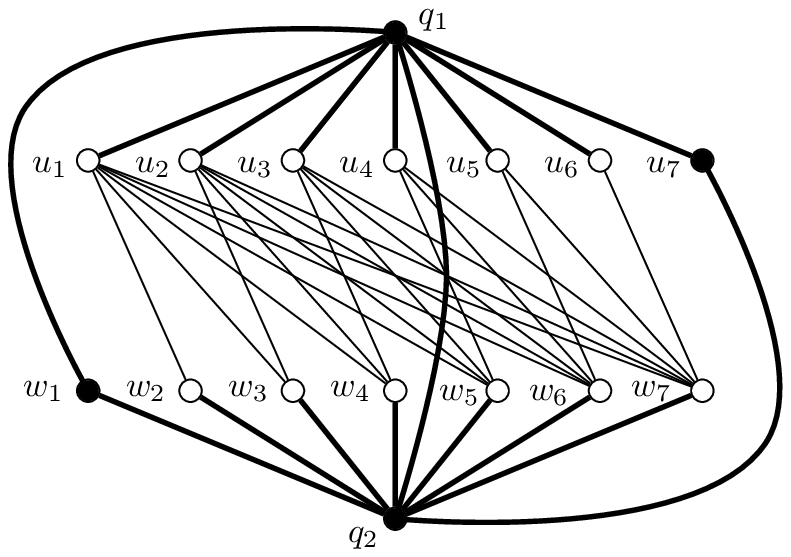}}
 \hspace{0.5cm}
\subfigure[Graph $L_2$]{\includegraphics[width=6.8cm,clip=true]{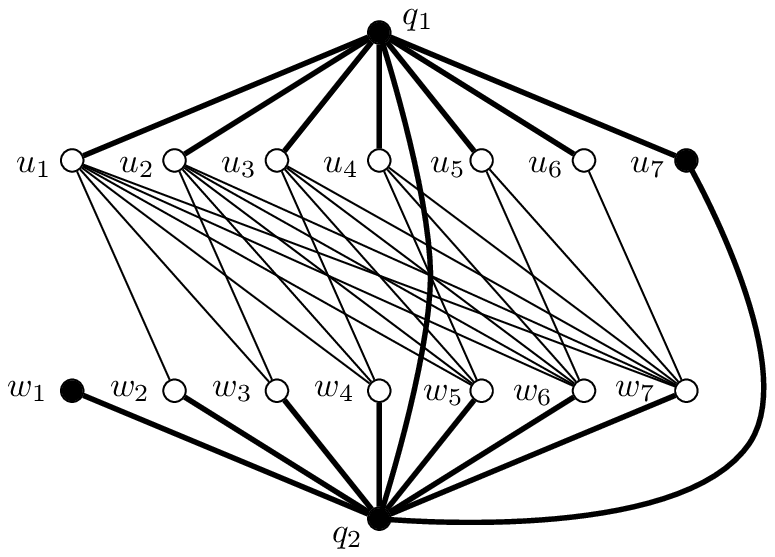}}
 \caption{Two extremal graphs}
 \label{L1L2}
\end{center}
\end{figure}

\msk
In Figure~\ref{L1L2} we present an example of $L_1$ and $L_2$, when $\ell=7$ (white vertices induce $B_{\ell}$).
From now on, we shall also refer to both graphs shortly as $L$.

The following claim contains an easier part of the proof of Lemma
\ref{ngeq}. Later we shall deal with a more difficult one.
\bcl \label{easyparts}
\be
  \item[\textup{(i)}]  \ $\chi (L)=3$,
  \item[\textup{(ii)}] \ $\psi (L)=h$.
\ee
\ecl

\begin{figure}[t]
\begin{center}
  \includegraphics[width=6.8cm,clip=true]{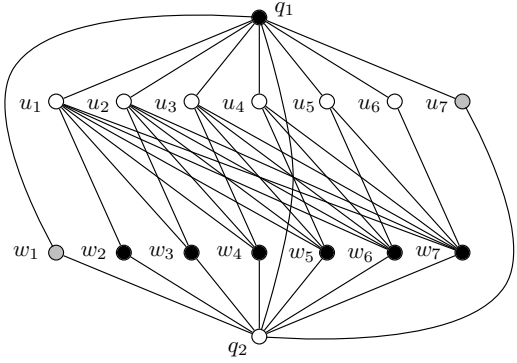}
  \caption{Proper 3-coloring of $L_1$}\label{colorfigure}
\end{center}
\end{figure}

\pf
\nin
(i) \ The graph $L$ contains a triangle, and we can easily find a coloring with
three colors, shown in Figure \ref{colorfigure}.

(ii) \ The vertex partition of $L$ into the $h$ stable sets
$\{q_1\}$, $\{q_2\}$, and $\{u_i,w_i\}$, $i\in\{1,\dots,\ell\}$ is a
complete coloring with exactly $h>3$ colors. Thus, $\psi(L) \geq h$.
Conversely, the upper bound $h$ for $\psi (L)$ comes from the
arguments used in the proof of {Lemma~\ref{fminus1}}.
The fact that $\vert V(L)\vert = 2h-2$ is also important but it follows easily from
the construction. \qed

\bsk
We continue with the harder part of the lemma. First we state a property similar to that of Claim \ref{minmax}.

\bcl \label{maxstab} The maximal stable sets\/ $S$ of graph\/
 $B_h'$ are of the following form:
$$
  S=S_N=\{w_1,\dots,w_N\} \cup \{u_N,\dots,u_{\ell}\}
$$
 for some $1 \leq N \leq {\ell}$. \qed
\ecl

We note that the extremal cases $N=1$ and $N=\ell$ correspond to
$U'\cup \{w_1\}$ and $W'\cup \{u_{\ell}\}$, respectively. Moreover
$\{u_{\ell},w_1\}$ is contained in all $S$, hence every $S$ meets
both $U'$ and $W'$.

\msk
Concerning stable sets of the graph $L$, we have the
following:

\bcl \label{Scompl} For any maximal stable set\/ $S$ of\/ $L$, the
induced subgraph\/ $L-S$ has property\/ $(\Pi)$ stated in Claim \ref{g=2}.
\ecl

\newpage

\pf Take an arbitrary $S$. There are three cases:

\be
\item[(a)] \ $\!S\subseteq V(B_h')$,

\item[(b)] \ $q_1\in S$,

\item[(c)] \ $q_2\in S$.
\ee

In the first case, $S$ is a maximal stable set of the graph $B_h'$,
and by Claim \ref{maxstab} we have $S=S_N$ for some $1\le N\le
\ell$. The complement of $S$ with respect to the vertex set of $L$
is the set $\{u_1,...,u_{N-1}\}\cup \{w_{N+1},...,w_{\ell}\}\cup
\{q_1,q_2\}$ and induces a complete bipartite graph with vertex
classes $\{u_1,...,u_{N-1}\}\cup \{q_2\}$ and
$\{w_{N+1},...,w_l\}\cup \{q_1\}$ since $u_{\ell}\in S$ necessarily
holds. This verifies property $(\Pi)$.

In the second case, $S\subseteq W'\cup \{q_1\}$ and, in fact, $S$ is
equal to this set, because of maximality. The complement of $S$ with
respect to the vertex set of $L$, namely $U'\cup \{q_2\}$, induces a
subgraph consisting of $U$ (see Construction I to recall the definition)
as $\ell-1$ isolated vertices, together with the isolated edge
$u_{\ell}q_2$. So it does have property $(\Pi)$.

The third case is similar and yields that $L-S$ is induced by the stable set
$W$ plus the isolated edge $u_{\ell}q_1$. \qed

\bsk

Finally, we prove

\bcl \label{three3} $\Gamma(L)=3$.
\ecl

\nin
\pf Let $\vp$ be an arbitrary Grundy coloring of $L$,
  and consider the
stable set $S$ formed by the vertices of color 1 under $\vp$. Since
every vertex of higher color has a neighbor of color 1, the set $S$ is a
maximal stable set of $L$. Thus, by Claim \ref{Scompl}, the
subgraph $L-S$ satisfies property $(\Pi)$. Now Claim
\ref{g=2} implies that every Grundy coloring of $L-S$ uses at most 2
colors. This implies $\Gamma (G)\leq 3$, whereas the presence of a subgraph
$K_3$ induced by $\{u_{\ell},q_1,q_2\}$ yields equality.
 \qed \bsk

To get a general construction of a graph that realizes $(f,g,h)$ with $2h-f+1$
vertices, it is enough to repeat the process of adding a clique of
order $f-3$, as we did at the end of the proof of Theorem
\ref{fkisebbg}. This completes the proof of Lemma \ref{ngeq}. \qed

\bsk

In this way Theorem \ref{f=g} has been proved as well.
\qed


\section{The lists of extremal graphs} \label{list}

Our goal in this section is to prove Theorem \ref{csak2}.
Along the proof we shall also describe the structure of extremal graphs that we call here $h$-optimal graphs.

\bdf Suppose $h\geq 4$. We say that a graph $G=(V,E)$ is \emph{$h$-optimal}
 if $\chi(G)=\Gamma(G)=3$, $\psi(G)=h$ and $\vert V(G)\vert=2h-2$.
\edf

In the following statements $G$ will be an $h$-optimal graph and
$\cal H$ will be an arbitrary, fixed complete $h$-coloring of $G$. Let us also recall that $\ell=h-2$.
The results in Section \ref{c=G} directly imply the following

\bcl \label{isClaim} In an\/ $h$-optimal graph, there are exactly
two color classes of one element and exactly\/ $\ell$ classes of
two elements in\/ $\cal H$. \qed \ecl

\bdf The color classes consisting of two elements will be called
\emph{pairs}. For a vertex $x$ in the pair, the other vertex will
be denoted mostly by $x'$; they will be the \emph{couples} of each
other. \edf

 \nin
{\bf Notation } We denote by $\phi _1$ and $\phi _2$ the two
vertices of the singleton color classes, and we set $\Phi = \{\phi _1,\phi_2\}$.
The pairs will be denoted by $M_1, M_2,\dots, M_{\ell}$ and $M$ will be their union.
Moreover, $F_i$ is the subgraph induced by the vertex set\/ $M_i \cup \Phi$.

 \bcl \label{P4K3}
 For any\/ $1\leq i \leq \ell$, the subgraph\/ $F_i$
 is isomorphic to a\/ $P_4$ or it contains a triangle as a subgraph. Consequently,\/
$\Gamma(F_i)=3$. \ecl

\pf
 Simply we use the fact that $\phi _1$ and $\phi _2$ are adjacent and because of the completeness of the
 coloring $\cal H$, both of them have at least one neighbor in $M_i$.
 The last statement follows from the monotonicity of $\Gamma$, and simple facts that $\Gamma(P_4)=\Gamma(K_3)=3$
 and that the only $4$-vertex graph having Grundy number greater than $3$ is $K_4$.
 \qed

\bdf If $F_i$ is isomorphic to $P_4$, then $M_i$ is called a \emph{pair of $P_4$-type},
     otherwise we call it a \emph{pair of $K_3$-type}.
\edf

 \nin
{\bf Notation } Let $S_i$ be any maximal stable set of $G$
containing $M_i$.

\bcl \label{1edge} For\/ $1\leq i,j\leq \ell$, the pairs\/ $M_i$ and\/ $M_j$
are joined by exactly one edge. \ecl

\pf  The completeness of $\cH$ implies that $S_i$ is disjoint from
$\Phi$. Furthermore, because of $\Gamma(F_j)= 3$, $S_i$ has to
intersect $F_j$ in some vertex, by Claim \ref{Gamma>}.
Consequently, $S_i \cap M_j \neq \emptyset$. Similarly we obtain
$S_j \cap M_i \neq \emptyset$. These conditions (two empty triples
inside $M_i\cup M_j$) leave room for just one edge between $M_i$
and $M_j$. \qed

\bcl \label{cliqueiso}
 If\/ $u \in M_i$ and\/  $\{\phi _1,\phi_2,u\}$ induces $K_3$,
 then\/ $u$ is an isolated vertex {in~$G[M]$}.
\ecl

\pf  Let $H$ be a triangle induced by $\{\phi _1,\phi_2,u\}$. Assume on the contrary that $u$ has a neighbor in $M_j$ for some $j \neq i$.
 Then the set $S_j$ is disjoint from $V(H)$ and dominates $H$, which implies $\Gamma (G)>3$,
 by Claim \ref{Gamma>}. A contradiction. \qed \bsk


From Claims \ref{P4K3} and \ref{cliqueiso} we obtain

\bc \label{ImplP4} If a pair has no isolated vertex in\/ $G[M]$,
then it is of\/ $P_4$-type. \qed \ec

\nin
 {\bf Definition and notation }
 Let $J$ be the union of the pairs containing some isolated vertex of the graph $G[M]$.
 In what follows we say that a vertex is \emph{non-isolated} if it is not isolated in $G[M]$.
 Let $T = M \setminus J$.

\msk
 In a series of the three subsequent claims we reveal the structure of the graph induced by $T$.
 Next, in Lemma \ref{twopairs} we analyze the number of pairs in the set $J$.

\bcl \label{Rbip}
    $T$ induces a bipartite graph.
\ecl

Before we prove Claim \ref{Rbip}, we give some definitions
and state some facts.

\bdf An induced $P_4$ of $G$ with the middle edge $\phi _1 \phi_2$ will
be called an \emph{emphasized $P_4$}. \edf

\bpr \label{P4endpoint} If a maximal stable set\/ $S$ contains
some pair then it intersects each emphasized\/ $P_4$ in at least
one endvertex. \epr

\pf As we know from Claim \ref{Gamma>}, the set $S$ intersects every
$P_4$. By assumption, $S$ contains some pair $M_i$. Using the
properties of the complete coloring, both $\phi _1$ and $\phi _2$
have some neighbor in $M_i$. Therefore $\phi _1, \phi _2 \not \in S$
and hence $S\cap P_4$ must be an endvertex. \qed

\msk
Now, let us define the following sets of vertices:

\begin{itemize}
\item[] $X=\{x \mid x\in T$ and $x$ is adjacent to $\phi _2\}$,
\item[] $Y=\{y \mid y\in T$ and $y$ is adjacent to $\phi _1\}$.
\end{itemize}

By Claim \ref{cliqueiso}, $T$ is the disjoint union of $X$ and $Y$, moreover, $|X|=|Y|$.

\bpr  \label{maxS} For any vertex\/ $x \in X$, there exists a
maximal stable set\/ $S$ containing some pair but not containing
$x$. The same is true for any\/ $y\in Y$, too. \epr

\pf By definition, every vertex in $T$ is non-isolated in $G[M]$.
Since $x \in T$, there exists a vertex $z$ adjacent to $x$ such that
$z \in M_i$ for some $i$. The set $S_i$ above can play the role of
$S$ in the proposition. \qed \bsk

 \nin
{\bf Proof of Claim \ref{Rbip} } We show that $Y$ does not induce
any edges. For $X$, the proof is analogous.

Suppose $\eta _1, \eta _2 \in Y$ and $\eta _1, \eta _2$ are adjacent. Then $\{ \eta _1 ', \phi _2, \phi _1, \eta _2\}$ induces a $P_4$ because $\eta _2$
has exactly one neighbor in $\{ \eta _1 ',\eta _1\}$. By Proposition \ref{maxS}, we have a maximal stable set $S$ containing some pair with $ \eta _1 '
\not \in S$. By Proposition \ref{P4endpoint}, $S$ intersects both induced 4-paths $ \eta _1 ' \phi _1 \phi _2 \eta _1$ and $ \eta _1 ' \phi_1 \phi _2 \eta
_2$, in one of their endvertices. It does not contain $ \eta _1 '$, thus it must contain both $\eta _1$ and $\eta_2$, a contradiction. \qed

\bcl \label{no2K2}
    $T$ induces a\/ $2K_2$-free graph.
\ecl

\pf  Assume on the contrary that we have a $2K_2$ in $G[T]$. We
denote its edges by $xy$ and $\overline{x}\,\overline{y}$.  Take the
6-vertex subgraph $H$ of $G$ induced by $\{x, \overline{x}, y, \overline{y}, \phi _1, \phi _2\}$.
The subgraph $H$ has a maximal stable set $\{x, \overline{y}\}$ and the remaining graph is a $P_4$,
because of the definition of the sets $X$ and $Y$. By Claim \ref{Gamma>}, we get $\Gamma(H)>3$, a contradiction. \qed

\msk Let $\tau$ be the number of pairs in $T$.
 It is a well-known fact that for a bipartite graph with partite sets
 $X,Y$ of the same cardinality, $2K_2$-freeness is equivalent to the following.

\bsk \nin \textbf{Property (\astx):}   $X$ and $Y$ can be ordered in such
a way that $X=(x_1, x_2, \dots, x_\tau)$, $Y=(y_1, y_2,\dots,y_{\tau})$,
and $N(x_i)\subseteq N(x_j)$ for every $i<j$ and $N(y_i)\supseteq N(y_j)$ for every $i<j$.

\bsk
 In the next claim we use the extended graphs $B'$ defined in Section~\ref{c=G}.

\bcl \label{Zsolt}
    The set\/ $T$ induces a graph isomorphic to the graph\/ $B_{\tau}'$.
\ecl

\pf  Let us pick a counterexample $T$ of smallest size. We state
that in $T$ the couple of $x_1$ is $y_1$. Suppose, for a
contradiction, that the couple of $x_1$ is $y_j$, for some $j>1$.
The pair $x_1 y_1$ cannot be an edge since otherwise, by Property
(\astx), $y_1$ would be adjacent to everything in $X$ and it would not
have any couple. From the nonadjacency of $x_1$ and $y_1$ and
Property (\astx), $x_1$ is isolated in $T$. Consequently, $y_j$ has
some neighbor in every class of the complete coloring, except its
own class $\{x_1, y_j\}$. Since $G[T]$ is bipartite, $y_j$ is
adjacent to all the vertices in $X\setminus x_1$. By Property (\astx), $y_1$ is
also adjacent to these vertices. Consequently, the only couple of
$y_1$ could be $x_1$, a contradiction. Thus, the couple of $x_1$
is $y_1$ indeed.

Taking the graph induced by $T' = T \setminus \{x_1, y_1\}$, would be a
smaller counterexample. Claim \ref{Zsolt} is proved.
\qed

\bsk
The next step is to manage the isolated vertices of $G[M]$.

 \bl  \label{twopairs}
  The set $J$ contains exactly two pairs.
 \el

 \pf
  Let $\xi$ be the number of pairs in $J$.

\bcl \label{isolleq} $\xi \leq 2$ \ecl

 \pf
  The assertion obviously holds for $\ell =2$. 
  Hence assume that $\ell\geq 3$.

\msk
 Suppose for a contradiction that $\xi \geq 3$.
 Without loss of generality, we may assume that $M_i=\{u_i,u'_i\}$,
 $i\in\{1,2,3\}$ are arbitrary pairs in $J$ such that $u_i$ is isolated in $G[M]$.
 By the completeness of the coloring ${\cal H}$, the non-isolated vertices $u'_i$ of pairs $M_i$ are mutually adjacent.
 In what follows we use $Q$ to denote the complete subgraph induced by $\{u'_1, u'_2, u'_3\}$.

 \msk
 If ${\ell} \geq 4$, then there exists a pair $M_j = \{r,r'\}$, $j > 3$.
 For each $i\in\{1,2,3\}$, consider the edge $e_i$ between $M_i$ and $M_j$.
 Obviously, $e_i$ contains $u_i'$. Consequently, the stable set $\{r,r'\}$ dominates $Q$.
 Using Claim \ref{Gamma>}, we obtain a contradiction. Hence it remains to consider the case when $\ell=3$.

 \msk
 Let $\ell=3$ and assume for a contradiction that $\xi \geq 3$, which in this case, by $\xi\leq\ell$, means $\xi = 3$.
 Also recall that under such assumptions we consider only $8$-vertex graphs.
 Now, observe that for $i\in\{1,2\}$ the vertex sets $N_i = V(Q) \setminus N(\phi _i)$ have the following properties:

\begin{itemize}
   \item[($\Pi_1$)] $N_1 \cup N_2=Q$.

Suppose not. Then, considering some uncovered vertex, it would be
isolated in $G[M]$ (by Claim \ref{cliqueiso}), contradicting the
completeness of the coloring ${\cal H}$. 
\item[($\Pi_2$)] $N_1 \cap N_2 \neq \emptyset$.

For a contradiction, using ($\Pi_1$), we may assume that $N_1= \{ u_2',u_3' \}$, $N_2 = \{ u_1' \}$.
Taking the triangle induced by $\{ \phi_2, u_2', u_3' \}$ and the stable set $M_1$ which dominates this triangle,
we get a contradiction by Claim \ref{Gamma>}.
\end{itemize}

Thus we may assume that $N_1$ and $N_2$ have some common vertex, say $u_1'$.
Hence, by the completeness of the coloring $\cal H$, it holds that $\phi _1 u_1 \in E$ and $\phi _2 u_1 \in E$.

Between the two sets $\{ \phi _1, \phi _2 \}$ and $\{ u_2', u_3' \}$
we have some edge, because of the connectedness condition, say $\phi
_2 u_2' \in E$. This implies $\phi _1 u_2' \notin E$, since
otherwise we would contradict ($\Pi_1$).

In order to avoid the $P_4$ induced by $\{ \phi _1, \phi _2, u_2', u_3' \}$ and dominated
by the stable set $M_1$, we state that, by ($\Pi_1$), either $\phi _1 u_3'$ or $\phi_2 u_3'$ is an edge.
If $\phi_2u_3'\in E$, then the triangle induced by $\{\phi_2, u_2', u_3'\}$
is dominated by $M_1$, a contradiction by Claim \ref{Gamma>}.
Hence $\phi_2 u_3'\notin E$, and consequently $\phi_1 u_3'\in E$.
However, in this case the $P_4$ induced by $\{ u_1, \phi _1, u_3', u_2' \}$ is dominated
by the stable set $\{ \phi _2, u_1'\}$. Since this cannot be affected by any further edges,
we get a contradiction by Claim \ref{Gamma>}.
%
\qed

\bcl \label{isolgeq} $\xi \geq 2$.
\ecl

\pf
 If $h=4$, then the assertion holds, since by Claim \ref{1edge}, each of the two pairs contains exactly one vertex that is isolated in $G[M]$.
 In what follows we assume that $h\geq 5$.

 \msk\noindent
 \textit{Case 1} We prove $\xi \neq 0$.

 \msk
 Suppose that $\xi=0$. By Claim \ref{Zsolt}, it holds that $G[T]$ is isomorphic to $B_{\tau}'$. There are two isolated
 vertices in this graph but $M = T$, by the assumption of the claim,
 which means that there is no isolated vertex in $M$, a contradiction.

 \msk\noindent
 \textit{Case 2} We prove $\xi \neq 1$.

  \msk
  Let $M_1=\{u_1,u'_1\}$ be a pair with $u_1$ being isolated in $G[M]$, and let $M_i=\{u_i,v_i\}$, $i\in\{2,\ldots,\ell\}$
  be the pairs that have no vertices that are isolated in $G[M]$, consequently, the pairs of $P_4$-type.
  Assume that $\{u_2,\ldots,u_\ell\}$ are adjacent to $\phi_1$, while $\{v_2,\ldots,v_\ell\}$ to $\phi_2$,
  and consider $M_1$, two distinct pairs $M_i, M_j$ chosen \emph{arbitrarily} from $M\setminus J$.
  Recall, that by Claim \ref{1edge} there are exactly three edges between the vertices of $M_1, M_i$ and $M_j$,
  while by Claim \ref{Zsolt} it holds that $u_i u_j, v_i v_j\notin E$.
  Consequently, we have three possibilities:
  \be
      \item[(a)] $u'_1v_i, u'_1u_j, u_iv_j \in E$,

      \item[(b)]  $u'_1v_i, u'_1u_j, v_iu_j \in E$,

      \item[(c)] $u'_1v_i, u'_1v_j, u_iv_j \in E$.
  \ee
  By symmetry, and by Claim \ref{Rbip}, there are no further possibilities.
  Also note, that besides the above-mentioned edges it will be enough to consider the edges between $\Phi$ and $M_1$
  and that by Claim \ref{cliqueiso}, the vertex $u'_1$ cannot be a common neighbor of $\phi_1$ and $\phi_2$.

  \msk
  We start with a simultaneous analysis of (a) and (b).
  Assume that $\phi_2u'_1\in E$.
  Then (a) implies that a path $P_4$ induced by $\{u_j,\phi_1,\phi_2,v_j\}$ is dominated by $\{u_i,u'_1\}$,
  while from (b) we obtain a triangle induced by $\{\phi_2,u'_1,v_i\}$ and dominated by $M_j$.
  Since this cannot be affected by adding any further edges, $\phi_2u'_1\notin E$ and hence, by completeness, $\phi_2u_1$ must be an edge.
  If so, then for (a), independently of whether $\phi_1u_1\in E$ or $\phi_1u'_1\in E$,
  a path $P_4$ induced by $\{u_j,\phi_1,\phi_2,v_i\}$ is dominated by $M_1$.
  Now, for (b), if $\phi_1u'_1\notin E$, then $\{u_1,u_j\}$ dominates a path $P_4$ induced by $\{u'_1,v_i,\phi_2,\phi_1\}$.
  On the other hand, if $\phi_1u'_1\in E$, then a triangle induced by $\{\phi_1,u_j,u'_1\}$ is dominated by $M_i$.
  Note that the analysis in case (b) is independent of whether $\phi_1u_1$ is an edge.

  Hence $\phi_2u_1\notin E$ and it finally follows that neither $\phi_2u'_1$ nor $\phi_2u_1$ is an edge,
  which contradicts the completeness of the coloring $\cal H$.

  It remains to consider case (c).
  First, observe that whenever all pairs $M_i, M_j$, $i,j\in\{2,\ldots,\ell\}$ satisfy $u'_1v_i, u'_1v_j, u_iv_j \in E$,
  then $u'_1$ is adjacent to each vertex in $\{v_2,\ldots,v_{\ell}\}$.
  Now, considering adjacencies between the pairs in $M\setminus J$, by Claim \ref{Zsolt}
  either $v_2, u_{\ell}$ or $v_{\ell},u_2$ are isolated in $G[M\setminus J]$.
  Extending the scope to $G[M]$, both $v_2$ and $v_{\ell}$ become neighbors of $u'_1$, but either $u_{\ell}$ or $u_2$
  remains isolated.
  This clearly contradicts our assumption that $\xi=1$.
  \qed

  \msk
  We have shown Claims \ref{isolleq} and \ref{isolgeq}, and thus Lemma \ref{twopairs} as well.
  \qed

\begin{figure}[t]
\begin{center}
  \includegraphics[width=0.95\textwidth,clip=true]{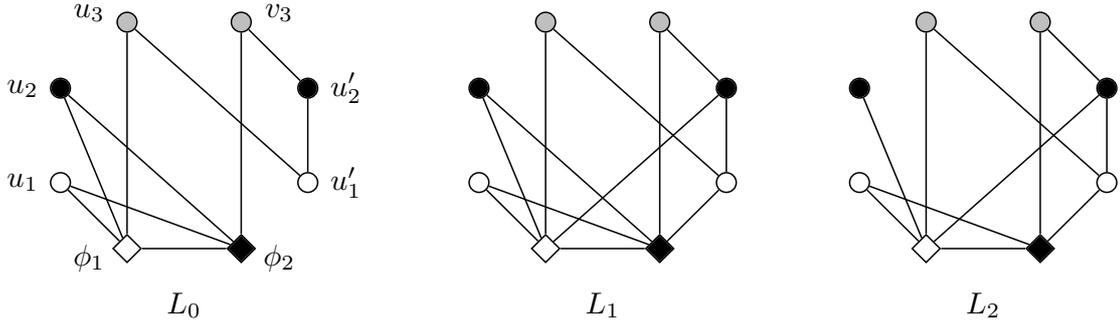}
  \caption{All $h$-optimal graphs for $h=5$ and their achromatic colorings ($\chi=\Gamma=3$, $\psi=5$, $n=2\psi-\chi+1=8$)}\label{fig:extremal335}
\end{center}
\end{figure}

\bsk
 Let $L_0,L_1$ and $L_2$ be the graphs presented in Figure \ref{fig:extremal335}.

 \bl \label{h5}
    If $h=5$, then a graph $G$ is $h$-optimal if and only if it is isomorphic to one of the graphs in $\{L_0,L_1,L_2\}$.
 \el

 \msk\pf
  By Lemma \ref{twopairs} a graph $G$ contains exactly one pair of $P_4$-type that consists of non-isolated vertices.
  Let $M_3=\{u_3,v_3\}$ be such a pair.
  For $i\in\{1,2\}$ let $M_i=\{u_i,u'_i\}$ be the pairs having $u_i$ isolated in $G[M]$.

 \msk
  Observe that $u'_1$ and $u'_2$ are adjacent and each of them must have some neighbor in $M_3$.
  Moreover, since vertices in $M_3$ are non-isolated, by Claim \ref{1edge} the neighbors of $u'_1,u'_2$ in $M_3$ must be disjoint.
  Without loss of generality we may assume that $u'_1u_3, u'_2v_3\in E$.
  By Claim \ref{1edge}, there are no other edges between $M_1,\ldots,M_3$, so it remains to consider the edges incident to $\phi_1$ or $\phi_2$.

 \msk
 Since $M_3$ is of $P_4$-type, assume that $\phi_1u_3,\phi_2v_3\in E$ and consequently $\phi_1v_3,\phi_2u_3\notin E$.
 Now, if both $\phi_2u'_1\in E$ and $\phi_2u'_2\in E$, then $M_3$ dominates a triangle induced by $\{\phi_2,u'_1,u'_2\}$, a contradiction.
 This cannot be affected by any further edges and hence $\phi_2$ cannot be simultaneously adjacent to $u'_1$ and $u'_2$.
 Consequently, by completeness, $\phi_2$ must be adjacent to at least one vertex in $\{u_1,u_2\}$.

 \msk\noindent
 \textit{Case 1} Assume that $\phi_2u'_1\in E$ and $\phi_2u'_2\notin E$.

 \msk
 Consequently, by completeness $\phi_2u_2\in E$.
 By Claim \ref{cliqueiso} we have $\phi_1u'_1\notin E$ and hence $\phi_1u_1\in E$.
 However, since the current graph is bipartite, we need to consider further edges.
 If $\phi_1u'_2\notin E$, then $M_3$ dominates a path $P_4$ induced by $\{\phi_1,\phi_2,u'_1,u'_2\}$,
 and this cannot be altered neither by $\phi_2u_1$ nor by $\phi_1u_2$.
 Hence $\phi_1u'_2\in E$. The graph is still bipartite.
 Now, adding either $\phi_2u_1$ or $\phi_1u_2$ results in the graph $L_1$, while adding both edges gives the graph $L_2$.

 \msk\noindent
 \textit{Case 2} Assume that $\phi_2u'_2\in E$ and $\phi_2u'_1\notin E$.

 \msk
  From Claim \ref{cliqueiso} it follows that $\phi_1u'_2\notin E$, while by completeness $\phi_1u_1\in E$ or $\phi_1u'_1\in E$.
  Assume that $\phi_1u'_1\in E$ and consider a subgraph $H$ induced by  $\Phi\cup M\setminus \{u_1,u_2\}$.
  Then a path $P_4$ induced by $\{\phi_1,u'_1,u'_2,v_3\}$ is dominated by $\{\phi_2,u_3\}$.
  Hence $\phi_1u'_1\notin E$.
  This in turn results in a graph containing a subgraph $P_4$ induced by $\{\phi_1,\phi_2,u'_1,u'_2\}$ and dominated by $M_3$.
  Since there are no further edges that could be added between vertices of $H$, we get a contradiction by Claim \ref{Gamma>}.

 \msk\noindent
 \textit{Case 3} Assume that $\phi_2u'_1,\phi_2u'_2\notin E$.

 \msk
  By completeness, $\phi_2u_1,\phi_2u_2\in E$. Note that it remains to consider the edges incident to $\phi_1$.
  Consider a subgraph $H$ induced by $\Phi\cup M \setminus \{u_1,u_2\}$.
  To argue that either $\phi_1$ is adjacent to both $u'_1$ and $u'_2$ or to none of them observe that whenever only one
  of the edges is present, then $M_3$ dominates a path $P_4$ induced by $\{\phi_1,\phi_2,u'_1,u'_2\}$.
  On the other hand, if we assume that both edges are present, then $M_3$ dominates a triangle induced by $\{\phi_1,u'_1,u'_2\}$.
  If both edges are missing, then by completeness $\phi_1u_1, \phi_2u_2\in E$, and we get the graph $L_0$,
  that obviously realizes a triple $(3,3,5)$.
    Thus, either we get a $h$-optimal graph $L_0$ or a contradiction by Claim \ref{Gamma>}.
 \qed



\begin{figure}[t]
\begin{center}
  \includegraphics[width=12cm,clip=true]{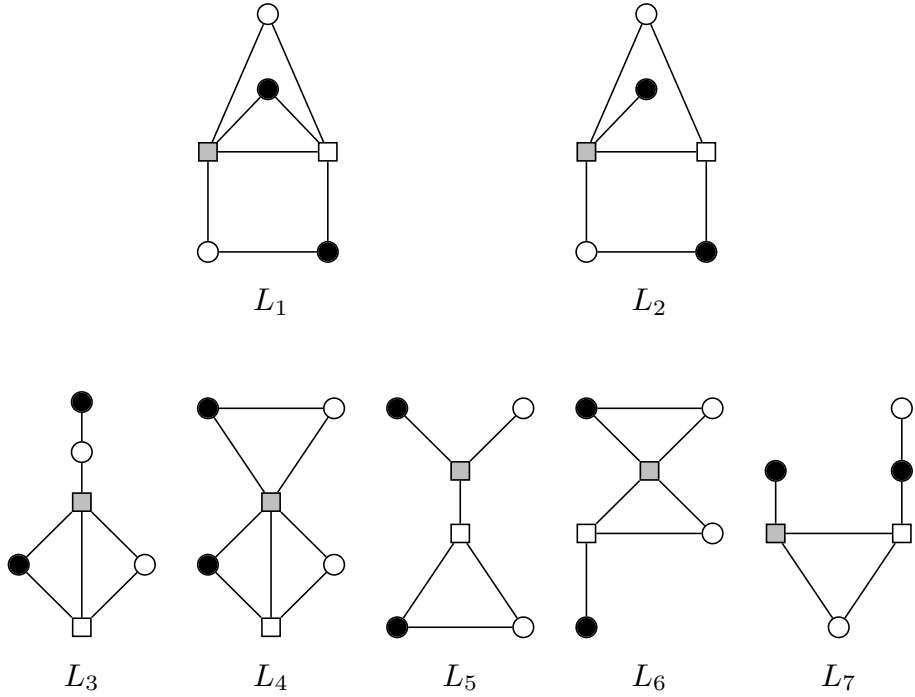}
  \caption{The seven $4$-optimal graphs and their achromatic colorings} \label{fig:extremal334}
\end{center}
\end{figure}

\bsk
 It is not hard to see that the arguments analogous to those in the proof of Lemma \ref{h5}
 can be used to obtain the list of all graphs that are $h$-optimal for $h=4$.
 Let $L_1,\ldots,L_7$ be the graphs presented in Figure \ref{fig:extremal334}.

\bl \label{h4}
    If $h=4$, then a graph $G$ is $h$-optimal if and only if it is isomorphic to one of the graphs in $\{L_1,\ldots,L_7\}$.
\el


\subsection{End of the proof of Theorem \ref{csak2} for $h\geq 6$}

By the analysis above, to  complete the proof of Theorem \ref{csak2}
 it remains to focus on $J$ and $\Phi$ when $h\geq 6$.

\msk
 \nin
{\bf Notation }
 Let $\iota_1, \iota_2$ be the vertices that are isolated in $G[M]$,
 and let $\nu_1, \nu_2$ be their couples, respectively.
Let $M_i =\{u_i,v_i\}$, $i\in\{1,\ldots,\tau\}$ be the pairs in $T$, that
is, the pairs without isolated vertices. Recall that $\tau=\ell-2=h-4$. 

\msk
The results on $2K_2$-free bipartite graphs entitle us to
suppose that the set of edges in $G[T]$ is $\{u_iv_I \mid I>i\}$.

\bcl \label{interval} $\{j \mid v_j\nu _1 \in E\}$ and $\{k \mid u_k\nu _1 \in
E\}$ are intervals. {\rm (}Here the empty set is also considered as an
interval.{\rm )} \ecl

\pf Suppose first that there are subscripts $j$ and $k>j$ such that
$\nu _1v_j \in E$ and $\nu _1u_k \in E$. Then we would have a stable
set $\{\iota _1,\nu _1\}$ dominating the $P_4$ induced by
$\{u_k,\phi _2,\phi _1,v_j\}$, a contradiction. (The relation of $\phi _1$ and
$\phi _2$ to the other vertices can be read out from the proof of
Claim \ref{Rbip}.)

 The vertex $\nu_1$ is adjacent to exactly one of $u_i$, $v_i$ for every $i$. Thus, denoting the neighbor of $\nu _1$ in $Y$ with the smallest
subscript by $v_j$, $N(\nu _1)\cap Y=\{v_j, v_{j+1},\ldots,v_{\tau}\}$ and $N(\nu _1)\cap X=\{u_1,u_2,\ldots,u_{j-1}\}$. \qed

\bcl \label{K4} One of $N(\nu _i)\cap X$, $N(\nu _i)\cap Y$
$(i\in\{1,2\})$ is empty. \ecl

\pf Otherwise $\{\nu _1,\nu _2,u_1,v_{\tau}\}$ would induce a $K_4$ and the Grundy number would be greater than 3. \qed

\bsk

We may assume $N(\nu _2)\cap X=\emptyset$. Then

\bcl \label{total}  {\rm (a)} $N(\nu _2)\cap Y=Y$, {\rm (b)} $N(\nu _1)\cap
Y=\emptyset$ and  {\rm (c)} $N(\nu _1)\cap X=X$. \ecl
\pf (a) is an obvious consequence of the assumption.

 Suppose $N(\nu _1)\cap Y$ is a nonempty proper subset of $Y$.
 If there was a $j$ such that $v_j\nu _1 \in E$ and $v_{j-1}\nu _1 \not \in E$, 
 then the stable set $\{u_{j-1}v_{j-1}\}$ would dominate the triangle induced by $\{\nu _1,\nu_2,v_j\}$, a contradiction. 
 Consequently, if the Claim was not true, then such a $j$ would not exist.

 If  $N(\nu _1) \cap Y = Y$ then one can find the stable set $\{u_1,v_1\}$
 that dominates the triangle induced by $\{\nu_1,\nu_2,v_2\}$, a contradiction again. (Note that we used $h \geq 6$.)

 Otherwise,  $N(\nu _1)\cap Y=\{v_1, v_2,\ldots,v_j\}$ for some $j<\tau$. 
 Thus $\nu _1v_{j+1}\not \in E$ and  $\nu_1u_{j+1} \in E$, contradicting the arguments
 in the proof of Claim \ref{interval}.
 This implies the validity of (b), from which (c) follows, too.
 \qed \bsk

 Now we can concentrate on the $8$ pairs of vertices between the sets
 $\{\phi _1,\phi _2\}$ and $\{\nu _1,\nu _2,\iota _1,\iota _2\}$
 since these are the remaining undetermined ones. First we prove

\bcl $\nu _2\phi _1 \not \in E$ (and $\nu _1\phi _2 \not \in E$). \ecl

\pf Otherwise the stable set $\{u_1,v_1\}$ would dominate the triangle induced by $\{\nu _2,\phi _1,v_{\tau}\}$. (The same works for the second statement.) \qed

\bcl   $\nu _1\phi _1\in E$ (and $\nu _2\phi _2 \in E$). \ecl

\pf Suppose $\nu _1 \phi _2 \not \in E$. Then the stable set $\{\nu _1,\phi _1\}$ dominates the $P_4$ induced by $\{v_1,\nu _2,v_2,u_1\}$. (The same works
for the second statement.)  \qed \bsk

The completeness of the coloring implies $\iota _2\phi _1\in E$ and $\iota _1\phi _2 \in E$. The last fact we need is

\bcl It is impossible that both $\iota _1\phi _1$ and $\iota _2\phi _2$ are non-edges. \ecl

\pf In this case the whole graph would be bipartite (with partite sets $X\cup \{\phi _1,\nu _2,\iota _1\}$ and $Y\cup \{\phi _2,\nu _1,\iota _2\}$), a
contradiction with the assumptions.
\qed \bsk

\msk Now we identify the notation above with that of the examples in
Figure \ref{L1L2} in the following way:
 $\phi_1 \longrightarrow q_2$, $\phi_2 \longrightarrow q_1$,
 $\iota_1 \longrightarrow u_{\tau+1}$,
 $\iota_2 \longrightarrow w_1$.


\msk

Let us look now at the graphs $L_1$ and $L_2$. In both graphs
$\iota _2 q _2 \in E$, moreover $\iota _2$ and $\phi _2$ are
adjacent in $L_1$, while they are non-adjacent in $L_2$. The only further
possible situation would be the converse but this would yield a
graph isomorphic to $L_2$, $L_1$ respectively.

\bsk

This completes the proof of Theorem \ref{csak2}. \qed

\section{Concluding remarks}

In this paper, for all realizable triples $(f,g,h)$ of integers,
 we determined the minimum order of connected graphs $G$
 such that $\chi(G)=f$, $\Gamma(G)=g$, and $\psi(G)=h$.
We completely described also the list of graphs attaining the
 minimum in all cases where $f<g\le h$ or $f=g=3$.
For the other triples the corresponding characterization of graphs
 remains unsolved:

 \bpm
   For larger common values \/ $f=g>3$, and\/ $h>f$, determine the list
   of\/ $h$-optimal graphs.
 \epm

Since the clique number is a universal lower bound on the
 chromatic number, one can study the extended chain of inequalities
  $\omega (G)\leq \chi (G)\leq \Gamma (G)\leq \psi (G)$.
In this context the following problem arises in a natural way:

 \bpm
   Let\/ $2\le a\le b\le c\le d$ be integers.
   \begin{itemize}
    \item [$(i)$]
     Give necessary and sufficient conditions for the existence
      of connected graphs\/ $G$ with\/
      $\omega (G)=a, \ \chi (G)=b, \ \Gamma (G)=c, \ \psi (G)=d$.
    \item [$(ii)$]
     If such graphs exist, determine their minimum order\/
      $n_0=n_0(a,b,c,d)$, and characterize the graphs whose number
      of vertices attains this minimum.
   \end{itemize}
 \epm

Probably, already some particular cases are quite hard:

 \bpm
  Solve the analogous problems for three-element subsets of\/
   $\{\omega, \, \chi, \, \Gamma, \, \psi\}$.
 \epm

Similar characterizations for graphs with restricted structural
 properties would also be of interest.


\begin{thebibliography}{99}

\bibitem{AHL}
 M. Ast\'e, F. Havet and C. Linhares-Sales,
 Grundy number and products of graphs.
 Discrete Math. 310 (2010), pp. 1482–-1490.

\bibitem{Bh}
 V. N. Bhave,
 On the pseudoachromatic number of a graph.
 Fund. Math. 102 (1979), pp. 159--164.

\bibitem{Bod}
 H. L. Bodlaender,
 Achromatic number is NP-complete for co-graphs and interval graphs.
 Inf. Proc. Lett. 31 (1989), pp. 135-–138.

\bibitem{online}
P. Borowiecki, On-line Coloring of Graphs.
 In: Graph Colorings (Kubale, ed.),
 Contemporary Mathematics, Vol. 352,
 Amer. Math. Soc. (2004), pp. 21--33.


\bibitem{BoRa13}
 P. Borowiecki, D. Rautenbach,
 New potential functions for greedy independence and coloring, submitted.

\bibitem{BoSi12}
 P. Borowiecki, E. Sidorowicz,
 Dynamic coloring of graphs.
 Fundamenta Informaticae 114 (2012), pp. 105--128.
 doi:10.3233/FI-2012-620

\bibitem{CaiEd}
 N. Cairnie, K. Edwards,
 Some results on the achromatic number.
 J. Graph Theory 26 (1997), pp. 129--136.

\bibitem{ChaOk}
 G. Chartrand, F. Okamoto, P. Zhang and Zs. Tuza,
 A note on graphs with prescribed complete coloring numbers.
 J. Combin. Math. Combin. Comput. LXXIII (2010), pp. 77--84.

\bibitem{Chau}
 A. Chaudhary,
 Approximation algorithms for the achromatic number.
 J. Algorithms 41 (2001), pp. 404--416.

\bibitem{ChrS79}
 C. A. Christen, S. M. Selkow,
 Some perfect coloring properties of graphs.
 J. Combin. Theory Ser. B 27 (1979), pp. 49--59.

\bibitem{GK}
 D. P. Geller, H. Kronk,
 Further results on the achromatic number.
 Fund. Math. 85 (1974), pp. 285--290.

\bibitem{Gru}
 P. M. Grundy,
 Mathematics and games.
 Eureka 2 (1939), pp. 6--8.

\bibitem{GyKi+97}
 A. Gy\'arf\'as, Z. Kir\'aly and J. Lehel,
 On-line 3-chromatic graphs - II. Critical graphs.
 Discrete Math. 177 (1997), pp. 99--122.

\bibitem{HKRS}
 M. M. Halld\'orsson, G. Kortsarz, J. Radhakrishnan, and S. Sivasubramanian,
 Complete partitions of graphs.
 Combinatorica 27 (2007), pp. 519–-550.

\bibitem{HHP}
 F. Harary, S. T. Hedetniemi and G. Prins,
 An interpolation theorem for graphical homomorphisms.
 Portugal. Math. 26 (1967), pp. 453--462.

\bibitem{HS10}
 F. Havet, L. Sampaio,
 On the Grundy number of a graph.
 In: Proc. Int. Symp. on Parameterized and Exact Computation (IPEC), December 2010,
 Lecture Notes in Computer Science, Vol.~6478, pp. 170--179.

\bibitem{HHB}
 S. M. Hedetniemi, S. T. Hedetniemi and T. Beyer,
 A linear algorithm for the grundy (coloring) number of a tree.
 Congr. Numer. 36 (1982), pp. 351-–363.

\bibitem{HM}
 P. Hell, D. J. Miller,
 Graphs with given achromatic number.
 Discrete Math. 16 (1976), pp. 195--207.

\bibitem{Kie98}
 H. A. Kierstead,
 Coloring graphs on-line. In: Online Algorithms --- The State of the Art (A. Fiat and G. J. Woeginger, eds.),
 Lecture Notes in Computer Science, Vol.~1442, Springer, Berlin, 1998, pp. 281--305.

\bibitem{Kor}
 G. Kortsarz,
 A lower bound for approximating the grundy number.
 Discrete Math. Theor. Computer Sci. 9 (2007), pp. 7-–22.

\bibitem{KorKra}
 G. Kortsarz, R. Krauthgamer,
 On the approximation of the achromatic number.
 SIAM J. Discrete Math. 14 (2000), pp. 408-–422.

\bibitem{KorShe}
 G. Kortsarz, S. Shende,
 Approximating the achromatic number problem on bipartite graphs.
 In: Proc. 11th European Symposium on Algorithms (Budapest, 2003),
 Lecture Notes in Computer Science, Vol.~2832, Springer, Berlin, 2003, pp. 385–-396.

\bibitem{ManMcD}
 D. Manlove, C. McDiarmid,
 The complexity of harmonious coloring for trees.
 Discrete Appl. Math. 57 (1995), pp. 133-–144.

\bibitem{TP}
 J. A. Telle, A. Proskurowski,
 Algorithms for vertex partitioning problems on partial k-trees.
 SIAM J. Discrete Math. 10 (1997), pp. 529--550.

\bibitem{YaGa}
 M. Yannakakis, F. Gavril,
 Edge dominating sets in graphs.
 SIAM J. Appl. Math. 38 (1980), pp. 364--372.

\bibitem{cobip}
 M. Zaker,
 Grundy chromatic number of the complement of bipartite graphs.
 Australas. J. Combin. 31 (2005), pp. 325--329.

\bibitem{Zak}
 M. Zaker,
 Inequalities for the Grundy chromatic number of graphs.
 Discrete Appl. Math. 155 (2007), pp. 2567--2572.

\bibitem{Zak2}
 M. Zaker,
 New bounds for the chromatic number of graphs.
 J. Graph Theory 58 (2008), pp. 110--122.


\end{thebibliography}
\end{document}